\documentclass[conference]{IEEEtran}
\usepackage{graphicx}
\usepackage{amssymb,amsmath}
\usepackage{verbatim}

\begin{document}

\title{Design of Pulse Shapes Based on Sampling with Gaussian Prefilter}
\author{
\IEEEauthorblockN{Edwin Hammerich}
\IEEEauthorblockA{Ministry of Defence, 95030 Hof, Germany\\
E-mail: edwin.hammerich@ieee.org}
}
\maketitle

\begin{abstract}
Two new pulse shapes for communications are presented. The first pulse shape generates a set of pulses 
without intersymbol interference (ISI) or ISI-free for short. In the neighborhood of the origin 
it is similar in shape to the classical cardinal sine function but is of exponential decay at infinity. 
This pulse shape is identical to the interpolating function of a generalized sampling theorem 
with Gaussian prefilter. The second pulse shape is obtained from the first pulse shape by spectral 
factorization. Besides being also of exponential decay at infinity, it has a causal appearance since it 
is of superexponential decay for negative times. 
It is closely related to the orthonormal generating function considered earlier by 
Unser in the context of shift-invariant spaces. This pulse shape is not ISI-free but it generates a 
set of orthonormal pulses. The second pulse shape may also be used to define a receive 
matched filter so that at the filter output the ISI-free pulses of the first kind are recovered.
\end{abstract}

\newtheorem{definition}{Definition}
\newtheorem{lemma}{Lemma}
\newtheorem{proposition}{Proposition}
\newtheorem{theorem}{Theorem}
\newtheorem{corollary}{Corollary}
\newtheorem{remark}{Remark}

\renewcommand{\d}{\mathrm{d}}
\newcommand{\e}{\mathrm{e}}
\renewcommand{\i}{\mathrm{i}}

\section{Introduction}
Unser \cite{Unser2000} extended the standard sampling paradigm to the representation or
even approximation of functions by elements of shift-invariant function spaces. These function spaces 
are defined by a generating function $\varphi$, which has to satisfy certain conditions. By one of them, 
the partition of unity condition, Gaussian functions are actually precluded from being used as generators. In 
\cite{Hammerich2} it was substantiated that Gaussian functions still could be useful 
generators. Consequently, following \cite[Tab.~1]{Unser2000}, the interpolating generating function and the 
dual generating function 
have been computed for a Gaussian generator in \cite{Hammerich2}. The computation of the corresponding 
orthonormal generating function, $\varphi_\mathrm{ortho}$, is accomplished in Section~\ref{SEC_SPECFAC} of the present paper. 
Rather than extracting the square root as suggested in \cite{Unser2000}, our approach is based on spectral 
factorization and leads, interestingly enough, to expressions in terms of $q$-analogs \cite{Andrews}. 
As an application, two new pulse shapes are proposed and discussed in Section~\ref{SEC_APP}.

The following notations and conventions are adopted: $L^2(\mathbb{R})$ is the space of square-integrable functions (or 
finite-energy signals) $f:\mathbb{R}\rightarrow\mathbb{C}\cup\{\infty\}$ with inner product 
$\langle f_1,f_2\rangle=\int_{-\infty}^\infty f_1(x)\overline{f_2(x)}\,\d x$ and norm 
$\|f\|=\langle f,f\rangle^{1/2}$. For the Fourier transform 
we use the definition $\hat{f}(\omega)=(2\pi)^{-1/2}\int_{-\infty}^\infty \e^{-\i\omega x}f(x)\,\d x$, 
where $x$ denotes time and $\omega$ angular frequency. $\ell^2(\mathbb{Z})$ is the space of square-summable, complex-valued 
sequences indexed by the integers. Finally, $\delta_n=0,\,n\in\mathbb{Z}\setminus\{0\}$, and 
$\delta_0=1$. 

\section{Sampling in Shift-Invariant Spaces and Localization Spaces Revisited}
The purpose of the present section is to motivate the pulse shapes presented in Section~\ref{SEC_APP}. 
To this end, we give a brief overview of sampling in shift-invariant spaces and localization spaces 
with an emphasis on those spaces defined by a Gaussian generator or a Gaussian prefilter respectively.

Suppose that $\varphi\in L^2(\mathbb{R})$ is a continuous function satisfying
$\varphi(x)=O(|x|^{-1-\epsilon})$ as $\,x\rightarrow\pm\infty$ for some $\epsilon>0$ and for any 
$\lambda>0$ the system of functions $\{\varphi(\cdot-n\lambda);n\in\mathbb{Z}\}$ forms a Riesz basis 
in $L^2(\mathbb{R})$. Furthermore, suppose that for any $\lambda>0$
\begin{equation}
  \sum_{n=-\infty}^\infty\varphi(n\lambda)\e^{-\i n\lambda\omega}\not=0,\;\omega\in\mathbb{R}. 
                                                                                     \label{walter_cdtn}
\end{equation}
The shift-invariant space $V_\lambda(\varphi)$ is the subspace of $L^2(\mathbb{R})$ defined as
\begin{equation}
   V_\lambda(\varphi)=\left\{f;f(x)=\sum_{n\in\mathbb{Z}} c_n\varphi(x-n\lambda),\,c\in
   \ell^2(\mathbb{Z})\right\}.                                                            \label{VSPACE}
\end{equation}
Then, the following sampling theorem applies.
\begin{theorem} \label{Theo1} For any $\lambda>0$ and $f\in V_\lambda(\varphi)$ it 
holds that
\begin{equation}
	f(x)=\sum_{n\in\mathbb{Z}} f(n\lambda)\varphi_{\mathrm {int}}(x-n\lambda),\;x\in\mathbb{R},
	                                                                            \label{interp_series}
\end{equation}
where the interpolating function $\varphi_{\mathrm {int}}\in V_\lambda(\varphi)$ is given by
\begin{align}
   \hat{\varphi}_\mathrm{int}(\omega)
     &=\frac{\hat{\varphi}(\omega)}{\sum_{n\in\mathbb{Z}}\varphi(n\lambda)\e^{-\i n\lambda\omega}} \nonumber\\
     &=\frac{\hat{\varphi}(\omega)}{\frac{\Lambda}{\sqrt{2\pi}}
              \sum_{k\in \mathbb{Z}}\hat{\varphi}(\omega+k\Lambda)},\quad\Lambda = \frac{2\pi}{\lambda}. \label{Lambda}					 					 
\end{align}
\end{theorem}

This theorem is due to Walter \cite{Walter}, who originally proved it for orthonormal bases 
$\{\varphi(\cdot-n);n\in\mathbb{Z}\}$. The theorem was later extended to Riesz bases by 
Unser~\cite{Unser2000}, who also considered alternative generating functions for $V_\lambda(\varphi)$ 
like $\varphi_\mathrm{int}$ as above and $\varphi_\mathrm{ortho}$ (see below). In an attempt to retain 
the flavor of the original Whittaker--Kotelnikov--Shannon (WKS) sampling theorem \cite{Jerri}, in 
\cite{Hammerich3} prior to sampling a prefilter
\begin{equation}
   (\boldsymbol{P}_\varphi f)(x)
     =\int_{-\infty}^\infty f(y)\overline{\varphi(y-x)}\,\d y \label{TP}
\end{equation}
with prefilter function $\varphi\in L^2(\mathbb{R})$ has been applied to an arbitrary finite-energy signal 
$f\in L^2(\mathbb{R})$. The so-called localization space
\[
  \mathcal{P}_\varphi =\{g=\boldsymbol{P}_\varphi f;f\in L^2(\mathbb{R})\}
\]
then corresponds to the space of bandlimited, finite-energy signals in the classical WKS sampling theorem. The 
goal is to recover the filter output signal $g=\boldsymbol{P}_\varphi f$ from sample values 
$g(n\lambda),\,n\in\mathbb{Z},$ either perfectly or at least with an acceptable error. To this end, the 
autocorrelation function $\Phi$ of $\varphi$,
\begin{equation}
   \Phi=\boldsymbol{P}_\varphi\varphi\in\mathcal{P}_\varphi 
                 \stackrel{\mathrm{Fourier}}{\longleftrightarrow}\hat{\Phi}(\omega)=
		           \sqrt{2\pi}|\hat{\varphi}(\omega)|^2,                   \label{Phi}
\end{equation}
is needed. The (second) interpolating function $\Phi_\mathrm{int}\in\mathcal{P}_\varphi $ is defined by its Fourier
transform
\begin{equation}
  \hat{\Phi}_\mathrm{int}(\omega)=\frac{\hat{\Phi}(\omega)}{\frac{\Lambda}{\sqrt{2\pi}}
                                        \sum_{k\in \mathbb{Z}}\hat{\Phi}(\omega+k\Lambda)}, \label{Phi_int}  
\end{equation}
where $\Lambda$ is as in \eqref{Lambda}. Note that because of the Riesz basis condition still imposed 
on $\varphi$ (see \cite{Hammerich3} for the full set of assumptions), which is equivalent to the 
existence of positive constants $A$ and $B$ (possibly depending on $\lambda$) so that \cite{AU1994}
\begin{equation}
  0<A\le \Lambda\!\sum_{k=-\infty}^\infty |\hat{\varphi}(\omega+k\Lambda)|^2\le B<\infty,
                                                                    \;\omega\in\mathbb{R},\label{Riesz}
\end{equation}
the denominator in \eqref{Phi_int} never will vanish. We remark that in general 
$\Phi_\mathrm{int}\neq \boldsymbol{P}_\varphi \varphi_\mathrm{int}$. In the special case of a Gaussian 
prefilter function,
\begin{equation}
	\varphi(x)\!=\!\frac{1}{\sqrt{2\pi}(1/\beta)}\,\e^{-\frac{x^2}{2(1/\beta)^2}}	
		\stackrel{\mathrm{Fourier}}{\longleftrightarrow}
		\hat{\varphi}(\omega)\!=\!\frac{1}{\sqrt{2\pi}}\,\e^{-\frac{\omega^2}{2\beta^2}},
		                                                                      \label{Gaussf}	
\end{equation}
where the parameter $\beta>0$ controls effective bandwidth, the following generalized sampling theorem 
has been obtained \cite{Hammerich2}, \cite{Hammerich3}.
\begin{theorem} \label{Theo2} For any $\lambda>0$ let the interpolating function $\Phi_{\mathrm{int}}$ be 
defined by (\ref{Phi_int}). Then for any function $g\in\mathcal{P}_\varphi$, where 
$g=\boldsymbol{P}_\varphi f,\,f\in L^2(\mathbb{R})$, the function $\tilde{g}$ given by
\begin{equation}
  \tilde{g}(x)=\sum_{n\in\mathbb{Z}} g(n\lambda)\Phi_\mathrm{int}(x-n\lambda),\;
                                                                        x\in \mathbb{R}, \label{reco}
\end{equation}
is again in $\mathcal{P}_\varphi$, it perfectly reconstructs $g$ at the sampling instants 
$x_n=n\lambda,\,n\in\mathbb{Z}$, and for all other $x\in \mathbb{R}$ the squared relative error 
$\left(|g(x)-\tilde{g}(x)|/\|f\|\right)^2$ becomes small as soon as $\lambda\le 1/\beta$, then decaying 
superexponentially to zero as $\lambda\rightarrow 0$.
\end{theorem}

For the Gaussian prefilter function $\varphi$ as given in \eqref{Gaussf} (which will be assumed for the rest of 
the paper) one has
\begin{equation}
  \hat{\Phi}_\mathrm{int}(\omega)=\frac{\lambda}{\sqrt{2\pi}}
          \frac{\e^{-\frac{\i\pi}{\tau}(\omega/\Lambda)^2}}
                {\sqrt{-\i\tau}\vartheta_3(\omega/\Lambda,\tau)}, \label{Phi_int_Gauss}
\end{equation}
where 
\begin{equation}
  \vartheta_3(z,\tau)=\frac{1}{\sqrt{-\i\tau}}\sum_{n=-\infty}^{\infty}
			                \e^{-\frac{\i\pi}{\tau}(z+n)^2}  \label{theta3}
\end{equation}
is a Jacobi theta function \cite{MOS} with parameter $\tau$ given by
\begin{equation}
  \tau=\i(\lambda\beta)^2\!/(4\pi). \label{tau}
\end{equation}
By inversion of the Fourier transform we obtain after use of Jacobi's $\tau\rightarrow -1/\tau$ transformation 
that \cite{Hammerich2}, \cite{Hammerich1}
\begin{equation}
  \Phi_{\mathrm{int}}(x)=
     \frac {\i\pi\tau}{\vartheta_1'\left(0,-1/\tau\right)}
	\frac{\vartheta_1\left(x/\lambda,-1/\tau\right)}{\sinh(\i\pi\tau x/\lambda)}, \label{interpol}
\end{equation}
where $\vartheta_1(z,\tau)=2\sum_{n=0}^{\infty}q^{\left(n+\frac{1}{2}\right)^2}(-1)^n\sin[(2n+1)\pi z],
\,q=\e^{\i\pi\tau},\,\Im(\tau)>0,$ is another theta function \cite{MOS}. When $\lambda\le 1/\beta$, it 
holds with high accuracy that $\Phi_{\mathrm{int}}(x)\approx S_0(x)$, where \cite{Hammerich1}
\begin{equation}
  S_0(x) \triangleq \i\tau\frac{\sin(\pi x/\lambda)}{\sinh(\i\pi\tau x/\lambda)},\;x\in\mathbb{R}. \label{S0}
\end{equation}
Note that in our context $\i\tau$ is always a negative real number and 
that $\Phi_{\mathrm{int}}(x)$ decays exponentially to zero as $x\rightarrow\pm\infty$.

\begin{figure}
\centering
\includegraphics[width=3.5in]{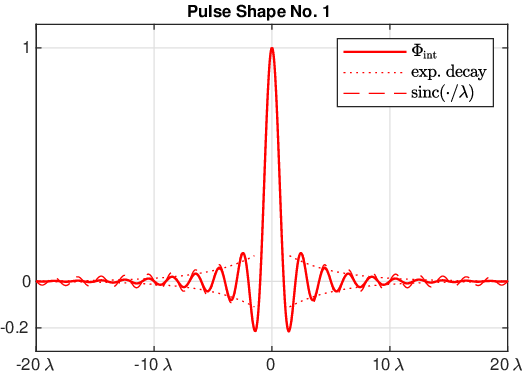}
\caption{$\Phi_\mathrm{int}$ for $\beta=100$ and $\lambda=1/\beta$; sinc is the cardinal sine function 
defined by $\mathrm{sinc}(x)=\sin(\pi x)/(\pi x)$.}
\label{Fig_1}
\end{figure}

Fig.~\ref{Fig_1} shows the interpolating function $\Phi_\mathrm{int}$ for $\beta=100$ and $\lambda=1/\beta$; actually, 
the approximation \eqref{S0} for $\Phi_\mathrm{int}(x)$ has been used.

\section{Spectral Factorization of the Interpolating Function} \label{SEC_SPECFAC}
We start by compiling a few prerequisites \cite{Andrews}.
\begin{definition} For any $a\in\mathbb{R}$ and $q\in\mathbb{C}$ with $|q|<1$ the $q$-Pochhammer symbol 
$(a;q)_n$ is defined by
\begin{align*}
  (a;q)_n&=(1-a)(1-aq)\cdots(1-aq^{n-1}),\,n=1,2,\ldots,\\
  (a;q)_\infty&=\lim_{n\rightarrow\infty}(a;q)_n\,,\\
  (a;q)_0&=1.
\end{align*}
\end{definition}
The following identities of Euler hold true for $q\in \mathbb{C},|q|<1$:
\begin{gather}
  1+\sum_{n=1}^\infty \frac{q^{\frac{1}{2}n(n-1)}}{(q;q)_n}z^n
                                =\prod_{n=0}^\infty(1+zq^n),\,z\in\mathbb{C}, \label{Euler1}\\
  1+\sum_{n=1}^\infty \frac{z^n}{(q;q)_n}
                                =\prod_{n=0}^\infty(1-zq^n)^{-1},\,z\in\mathbb{C},|z|<1. \label{Euler2}								
\end{gather}
Only Eq.~\eqref{Euler1} will be needed in the proof of the next theorem, where Jacobi's triple product identity
\begin{equation}
  \sum_{n=-\infty}^\infty q^{n^2}z^n
      =\prod_{n=1}^\infty(1-q^{2n})(1+q^{2n-1}z^{-1})(1+q^{2n-1}z)  \label{Jacobi}
\end{equation}
will play a central role. In our case always
\begin{equation}
   q=\e^{\i\pi\tau},   \label{q}
\end{equation}
where parameter $\tau$ is as in \eqref{tau}. The special $q$-Pochhammer symbols
\[
   (q^2;q^2)_n=\prod_{k=1}^n(1-q^{2k}),\quad Q_0\triangleq(q^2;q^2)_\infty
\]
will occur frequently.
\begin{theorem} \label{Theo3} For the function
\begin{equation}
  \varphi_\mathrm{ortho}(x)=\frac{1}{\sqrt{(q^2;q^2)_\infty}}
                 \sum_{n=0}^\infty\frac{(-1)^n q^n}{(q^2;q^2)_n}\,\phi(x-n\lambda),   \label{varphi_ortho}
\end{equation}
where $x\in\mathbb{R}$ and
\begin{equation}
  \phi(x)=\frac{\beta^{1/2}}{\pi^{1/4}}\,\e^{-\frac{x^2}{2(1/\beta)^2}}     \label{phi}
\end{equation}
is the Gaussian function $\varphi$ in \eqref{Gaussf} normalized to unit energy, i.e., 
$\int_\mathbb{R} |\phi(x)|^2\,\d x=1$, it holds that 
\begin{equation}
  \hat{\Phi}_\mathrm{int}(\omega)=\sqrt{2\pi}|\hat{\varphi}_\mathrm{ortho}(\omega)|^2,\;\omega\in\mathbb{R}.
                                                                                               \label{Bdg2}
\end{equation}
\end{theorem}
\begin{IEEEproof} The definition (\ref{Phi_int_Gauss}) of the interpolating function 
$\Phi_\mathrm{int}$ may be written in the Fourier domain as
\[
  \hat{\Phi}_\mathrm{int}(\omega)=\sqrt{2\pi}\lambda\frac{|\hat{\varphi}(\omega)|^2}
                                         {\sqrt{-\i\tau}\vartheta_3(\omega/\Lambda,\tau)}.
\]
Then, Eq. (\ref{Bdg2}) becomes
\[
  |\hat{\varphi}_\mathrm{ortho}(\omega)|^2=
    \lambda\frac{|\hat{\varphi}(\omega)|^2}{\sqrt{-\i\tau}\vartheta_3(\omega/\Lambda,\tau)}.
\]
Since the theta function (\ref{theta3}) has the second representation \cite{MOS}
\[
  \vartheta_3(x,\tau)=1+2\sum_{n=0}^\infty q^{n^2}\cos(2\pi nx)
                     =\sum_{n=-\infty}^\infty q^{n^2}\e^{2\pi\i nx} \label{theta3a},
\]
we obtain by means of Eq.~(\ref{Jacobi}), putting
\[
     P(z)=\prod_{n=1}^\infty(1+q^{2n-1}z^{-1})                                                                                
\]
and subsequently $z=\e^{2\pi\i x}$, for $\vartheta_3(x,\tau)$ the factorization
\[
  \vartheta_3(x,\tau)=Q_0^{1/2}P(\e^{2\pi\i x})
	                       \cdot\overline{Q_0^{1/2}P(\e^{2\pi\i x})},\; x\in\mathbb{R}.
\]
Since the function $x\mapsto\vartheta_3(x,\tau)$ is real-valued and positively lower bounded on 
$\mathbb{R}$ so is the function $x\mapsto |P(\e^{2\pi\i x})|$. As a consequence, the definition of 
$\varphi_\mathrm{ortho}$ in the Fourier domain by 
\[
   \hat{\varphi}_\mathrm{ortho}(\omega)=\lambda^\frac{1}{2}\frac{\hat{\varphi}(\omega)}
                                  {(-\i\tau)^{1/4}Q_0^{1/2}P(\e^{2\pi i\omega/\Lambda})}
\]
will result in a function $\varphi_\mathrm{ortho}\in L^2(\mathbb{R})$ satisfying Eq.~\eqref{Bdg2}.

Now, we need to invert the Fourier transform. Since $x\mapsto 1/P(\e^{2\pi\i x})$ is a bounded 
1-periodic function, it is in $L^2([0,1))$ and thus has a Fourier series expansion 
$1/P(\e^{2\pi\i x})=\sum_{n=-\infty}^\infty a_n\e^{-2\pi\i nx},\,
a\in\ell^2(\mathbb{Z}),$ with the coefficients
\begin{equation}
  a_n=\int_0^1\e^{2\pi\i nx}\frac{1}{P(\e^{2\pi\i x})}\,\d x,\,n\in\mathbb{Z}.  \label{a_n}
\end{equation}
By inverse Fourier transform we now obtain (putting $c=\lambda^{1/2}(-\i\tau)^{-1/4}Q_0^{-1/2}$) 
that
\begin{eqnarray*}
  \varphi_\mathrm{ortho}(x)&=&\frac{c}{\sqrt{2\pi}}\int_{-\infty}^\infty \e^{\i x\omega}
            \frac{\hat{\varphi}(\omega)}{P(\e^{2\pi\i\omega/\Lambda})}\,\d\omega \\
         &=&c\sum_{n=-\infty}^\infty\frac{a_n}{\sqrt{2\pi}}\int_{-\infty}^\infty \e^{\i(x-n\lambda)\omega}
	                                                           \hat{\varphi}(\omega)\,\d\omega\\
	 &=&c\sum_{n=-\infty}^\infty a_n\varphi(x-n\lambda)\\
	 &=&\frac{1}{Q_0^{1/2}}\sum_{n=-\infty}^\infty a_n\phi(x-n\lambda).
\end{eqnarray*}

The computation of the coefficients (\ref{a_n}) is carried out in the complex domain.

\textbf{Case} $n=-1,-2,\ldots:$ After substitution in \eqref{a_n} of $x$ by $1-x$ 
we obtain
\begin{eqnarray*}
  a_n&=&\int_0^1\e^{-2\pi\i nx}\frac{1}{P(\e^{-2\pi\i x})}\,\d x \\
     &=&\frac{1}{2\pi\i}\oint_{|z|=1}\frac{z^{-n-1}}{P(z^{-1})}\,\d z, \\
\end{eqnarray*}
where integration in the contour integral is performed counterclockwise around the unit circle. Since the 
integrand function is analytic within a neighborhood of the closed unit disc, we obtain by means of 
Cauchy's integral theorem that $a_n=0,\,n=-1,-2,\ldots$

\textbf{Case} $n=0,1,\ldots:$ Eq.~\eqref{a_n} now directly yields by transition to 
a contour integral that
\begin{eqnarray*}
  a_n &=&\frac{1}{2\pi\i}\oint_{|z|=1}\frac{z^{n-1}}{P(z)}\,\d z \\
      &=&\lim_{M\rightarrow\infty}\underbrace{\frac{1}{2\pi\i}\oint_{|z|=1}\frac{z^{n-1}}{P_M(z)}\,\d z}_
                                                                                                {I_M(n)},
\end{eqnarray*}
where
\[
  P_M(z)=\prod_{k=0}^{M-1}(1+q^{2k+1}z^{-1}),\,M=1,2,\ldots,
\]
and the path of integration is same as before. The integrand function in the integral defining $I_M(n)$ 
has simple poles at
\[
   z_m = -q^{2m+1},\,m=0,1,\ldots,M-1,
\]
lying inside of the unit circle. By means of the theorem of residues we obtain
\[
  I_M(n) = \sum_{m=0}^{M-1} \mathrm{Res}_{z_m}\frac{z^{n-1}}{P_M(z)},
\]
whence
\[
  a_n=\lim_{M\rightarrow\infty}I_M(n)=\sum_{m=0}^\infty \mathrm{Res}_{z_m}\frac{z^{n-1}}{P(z)}.
\]
We compute that
\begin{eqnarray*}
  \mathrm{Res}_{z_m}\frac{z^{n-1}}{P(z)}&=&\mathrm{Res}_{z_m}\frac{z^n}{zP(z)} \\
    &=&\mathrm{Res}_{z_m}\frac{z^n}{(z-z_m)\prod_{k=0, k \ne m}^\infty (1+q^{2k+1}z^{-1})} \\
    &=&\frac{z_m^n}{\prod_{k=0, k \ne m}^\infty (1+q^{2k+1}z_m^{-1})} \\
    &=&\frac{(-q)^n q^{2mn}}{(q^2;q^2)_\infty\prod_{k=1}^m(1-q^{-2k})} \\
    &=&\frac{(-q)^n}{(q^2;q^2)_\infty}\,\frac{ (-1)^mq^{m(m+1)}(q^{2n})^m}{\prod_{k=1}^m(1-q^{2k})},
\end{eqnarray*}
treating in the case of $m=0$ the empty product as one. After replacement of $q$ in Eq.~\eqref{Euler1} 
with $q^2$ we get
\[
  1+\sum_{m=1}^\infty \frac{q^{m(m-1)}}{(q^2;q^2)_m}z^m=\prod_{m=0}^\infty(1+zq^{2m}).
\]
Hence it holds that
\begin{eqnarray*}
  a_n &=& \frac{(-q)^n}{(q^2;q^2)_\infty}
        \sum_{m=0}^\infty \frac{ q^{m(m-1)}}{(q^2;q^2)_m}(-q^{2(n+1)})^m\\
      &=& \frac{(-q)^n}{(q^2;q^2)_\infty}\prod_{m=0}^\infty(1-q^{2(n+1)}q^{2m})\\
      &=& \frac{(-q)^n}{(q^2;q^2)_n},	
\end{eqnarray*}
which concludes the proof of Theorem~\ref{Theo3}.
\end{IEEEproof}

\begin{figure}
\centering
\includegraphics[width=3.5in]{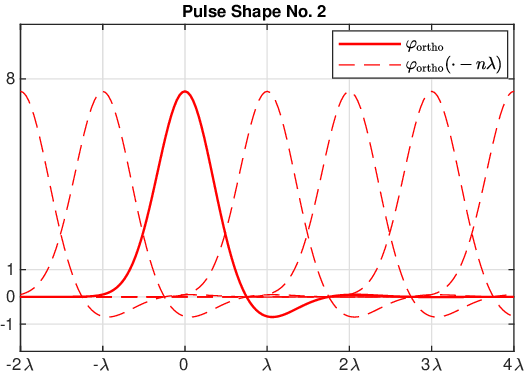}
\caption{$\varphi_\mathrm{ortho}$ together with some of its orthogonal translates for $\beta=100$ and 
$\lambda=3/\beta$.}
\label{Fig_2}
\end{figure}

In Fig.~\ref{Fig_2}, the function $\varphi_\mathrm{ortho}$ along with some of its translates 
is depicted for bandwidth parameter $\beta=100$ and $\lambda=3/\beta$. 
\begin{proposition} \label{Propo}
The function $\phi$ in \eqref{phi} has for $x\in\mathbb{R}$ the representation
\begin{equation}
  \phi(x)=Q_0^{1/2}\sum_{n=0}^\infty\frac{q^{n^2}}{{(q^2;q^2)_n}}\,\varphi_\mathrm{ortho}(x-n\lambda), \label{repr_phi}
\end{equation}
where $q$ is as in Theorem~\ref{Theo3}
\end{proposition}
\begin{IEEEproof}
The function $\varphi_\mathrm{ortho}$ as given in Eq.~\eqref{varphi_ortho} has the Fourier transform
\begin{align*}
  \hat{\varphi}_\mathrm{ortho}(\omega) &= \frac{1}{\sqrt{2\pi}}\int_{-\infty}^\infty \e^{-\i\omega x}\varphi_\mathrm{ortho}(x)\,\d x \\
     &=\frac{1}{Q_0^{1/2}}\sum_{n=0}^\infty\frac{(-q)^n}{(q^2;q^2)_n}\,\e^{-\i n\lambda\omega}\hat{\phi}(\omega) \\
     &=\frac{1}{Q_0^{1/2}}\sum_{n=0}^\infty\frac{(-q\e^{-\i\lambda\omega})^n}{(q^2;q^2)_n}\,\hat{\phi}(\omega).
\end{align*}
Because of Eq.~\eqref{Euler2} it holds that
\[
  \sum_{n=0}^\infty\frac{(-q\e^{-\i\lambda\omega})^n}{(q^2;q^2)_n}=\prod_{n=0}^\infty(1+q\e^{-\i\lambda\omega}\cdot q^{2n})^{-1}.
\]
We thus obtain
\[
  \hat{\varphi}_\mathrm{ortho}=\frac{1}{Q_0^{1/2}}\prod_{n=0}^\infty[1+q\e^{-\i\lambda\omega}(q^2)^n]^{-1}\hat{\phi}(\omega),
\]
which is equivalent to
\begin{equation}
  \hat{\phi}(\omega)=Q_0^{1/2}\prod_{n=0}^\infty[1+q\e^{-\i\lambda\omega}(q^2)^n]\,\hat{\varphi}_\mathrm{ortho}(\omega). \label{Eq_hatphi}
\end{equation}
Because of Eq.~\eqref{Euler1} it furthermore holds that
\begin{align*}
  \prod_{n=0}^\infty[1+q\e^{-\i\lambda\omega}(q^2)^n]&=\sum_{n=0}^\infty \frac{q^{n(n-1)}}{(q^2;q^2)_n}(q\e^{-\i\lambda\omega})^n \\
     &=\sum_{n=0}^\infty \frac{q^{n^2}}{(q^2;q^2)_n}\,\e^{-\i n\lambda\omega}.
\end{align*}
Inserting the last expression into Eq.~\eqref{Eq_hatphi} and inverting the Fourier transform yields
\begin{align*}
  \phi(x)&=Q_0^{1/2}\sum_{n=0}^\infty\frac{q^{n^2}}{(q^2;q^2)_n}\frac{1}{\sqrt{2\pi}}
                   \int_{-\infty}^\infty\e^{\i\omega(x-n\lambda)}\hat{\varphi}_\mathrm{ortho}(\omega)\,\d\omega \\
         &=Q_0^{1/2}\sum_{n=0}^\infty\frac{q^{n^2}}{{(q^2;q^2)_n}}\,\varphi_\mathrm{ortho}(x-n\lambda),
\end{align*}
which concludes the proof of Proposition~\ref{Propo}.
\end{IEEEproof}
\begin{corollary} \label{Coro}
Let the Gaussian generator $\varphi$ be as given in \eqref{Gaussf} and let the function $\varphi_\mathrm{ortho}$ be as given in 
Eq.~\eqref{varphi_ortho}. Then for any $\lambda>0$ it holds that
\begin{equation}
  V_\lambda(\varphi_\mathrm{ortho})=V_\lambda(\varphi).  \label{coro}
\end{equation}
\end{corollary}
\begin{IEEEproof}
For any generating function $\varphi\in L^2(\mathbb{R})$ and any $\lambda>0$, the shift-invariant space $V_\lambda(\varphi)$ as given in 
\eqref{VSPACE} may also be defined as the closed linear span of the system of functions $\{\varphi(\cdot-n\lambda);\,n\in\mathbb{Z}\}$ in 
$L^2(\mathbb{R})$; in particular, $V_\lambda(\varphi)$ is a closed linear subspace of $L^2(\mathbb{R})$. In the case of the above Gaussian 
generator $\varphi$, due to Theorem~\ref{Theo3}, $\varphi_\mathrm{ortho}\in V_\lambda(\phi)=V_\lambda(\varphi)$ so that
\begin{equation}
  V_\lambda(\varphi_\mathrm{ortho})\subseteq V_\lambda(\varphi).   \label{Ineq1}
\end{equation}
On the other hand, by reason of Proposition~\ref{Propo}, $\varphi\propto\phi\in V_\lambda(\varphi_\mathrm{ortho})$ so that
\begin{equation}
  V_\lambda(\varphi)\subseteq V_\lambda(\varphi_\mathrm{ortho}).   \label{Ineq2}
\end{equation}
The set inclusion \eqref{Ineq1} in combination with the set inclusion \eqref{Ineq2} results in the set equality \eqref{coro}, which concludes 
the proof of Corollary~\ref{Coro}.
\end{IEEEproof}

\section{Applications} \label{SEC_APP}

\subsection{ISI-Free Pulses} \label{SEC_ISIFREE}

The function $\Phi_\mathrm{int}$ in \eqref{interpol} may be used as a pulse shape to generate ISI-free 
pulses. Indeed, from representation \eqref{Phi_int} it is readily seen that for all $\omega\in\mathbb{R}$ it 
holds
\begin{equation}
  \sum_{n\in\mathbb{Z}} \Phi_\mathrm{int}(n\lambda)\e^{-\i n\lambda\omega}=
     \frac{\Lambda}{\sqrt{2\pi}}\sum_{k\in\mathbb{Z}} \hat{\Phi}_\mathrm{int}(\omega+k\Lambda)=1
                                                                                      \label{Bdg1}
\end{equation}
(the first equation being Poisson's summation formula; e.g., \cite{Unser2000}), 
which is equivalent to $\Phi_\mathrm{int}(n\lambda)=\delta_n,\,n\in\mathbb{Z}$. 
Therefore, the set of shifted pulses $\{\Phi_\mathrm{int}(x-n\lambda);n\in\mathbb{Z}\}$ is 
ISI-free at points in time $x_n=n\lambda$. The second equation in \eqref{Bdg1} is, of course, an instance of the 
well-known Nyquist criterion for ISI-free pulses \cite{Proakis}. Concerning the use of the also ISI-free 
pulses generated by \eqref{S0} (for \textit{arbitrary} positive parameters $\beta$ and $\lambda$) refer to 
\cite{KraftZoelzer}.

\subsection{Orthonormal Pulses with ISI-Free Matched Filter Output}

\subsubsection{Orthonormal Pulses} 
For any function $\varphi\in L^2(\mathbb{R})$ the system of functions 
$\{\varphi(x-n\lambda);n\in \mathbb{Z}\}$ forms an orthonormal system in $L^2(\mathbb{R})$ if and only 
if in \eqref{Riesz} $A=B=1$ may be chosen \cite{Unser2000}. For the 
function $\varphi_\mathrm{ortho}$ of Theorem~\ref{Theo3} this is true because of Eq.~\eqref{Bdg2} 
and the second equation in \eqref{Bdg1}. Therefore, $\varphi_\mathrm{ortho}$ may be used as a pulse 
shape to generate a set $\{\varphi_\mathrm{ortho}(x-n\lambda);\,n\in\mathbb{Z}\}$ of orthonormal pulses.

\subsubsection{Matched Filter} 
At the receiver, the filter $\boldsymbol{P}_{\varphi_\mathrm{ortho}}$ obtained by replacing the 
prefilter function $\varphi$ in \eqref{TP} with $\varphi_\mathrm{ortho}$ forms a matched filter 
allowing optimal detection of the pulses $\varphi_\mathrm{ortho}(x-n\lambda)$ at 
points in time $x_n=n\lambda,\;n\in\mathbb{Z},$ in the presence of noise. Due to orthogonality, 
overlap of adjacent pulses is of no relevance. Moreover, since it holds that [cf. \eqref{Phi} and Eq.~\eqref{Bdg2}]
\newpage
\[
  \boldsymbol{P}_{\varphi_\mathrm{ortho}}\varphi_\mathrm{ortho}=\Phi_\mathrm{int},
\]
the ISI-free pulses $\Phi_\mathrm{int}(x-n\lambda),\,n\in\mathbb{Z},$ of Section~\ref{SEC_ISIFREE} 
are recovered at the matched filter output.

In the light of the last application, the two proposed pulse shapes $\Phi_\mathrm{int}$ and 
$\varphi_\mathrm{ortho}$ are seen to correspond to the classical pulse shapes produced by a 
raised-cosine filter or a root raised-cosine filter respectively \cite{Proakis}. Recall that 
the proposed pulse shapes decay (super-)exponentially to zero as $x\rightarrow\pm\infty$ whereas the 
classical pulse shapes merely decay order of $x^{-3}$.

%

\section*{acknowledgement}
The author wishes to thank Ayush Bhandari for interesting discussions and helpful suggestions concerning 
an earlier draft of this paper.

\end{document}